\documentclass[apj]{emulateapj}
\usepackage{apjfonts}
\usepackage[usenames,dvipsnames]{xcolor}
\usepackage{slantsc}
\usepackage[T1]{fontenc}
\definecolor{blue}{rgb}{0,0,1}
\usepackage{CJK}
\usepackage{float}
\usepackage{multirow}
\usepackage{color}

\newcommand{\bdv}[1]{\mbox{\boldmath$#1$}}
\def\masyr{{\rm mas~yr^{-1}}}
\usepackage[hyperfootnotes=false]{hyperref}
\hypersetup{
  colorlinks,
  citecolor=blue,
  linkcolor=blue,
  urlcolor=blue}

\slugcomment{}
\shortauthors{Zhu et al.}
\begin{document}

\title{Mass Measurements of Isolated Objects from Space-based Microlensing}
\begin{CJK*}{UTF8}{gkai}

\author{Wei~Zhu~(祝伟)\altaffilmark{1,*},
S.~Calchi~Novati$^{2,3,4,a}$, 
A.~Gould\altaffilmark{1},
A.~Udalski\altaffilmark{5}, 
C.~Han\altaffilmark{6},  
Y.~Shvartzvald$^{7,8,b}$, 
C.~Ranc\altaffilmark{9}, 
U.~G.~J{\o}rgensen\altaffilmark{10}, 
R.~Poleski\altaffilmark{1,5},
V.~Bozza\altaffilmark{3,11} \\
and \\
C.~Beichman\altaffilmark{2}, 
G.~Bryden\altaffilmark{7},
S.~Carey\altaffilmark{2},
B.~S.~Gaudi\altaffilmark{1},
C.~B.~Henderson$^{1,7,b}$,
R.~W.~Pogge\altaffilmark{1},
I.~Porritt\altaffilmark{12},
B.~Wibking\altaffilmark{1},
J.~C.~Yee$^{13,c}$ \\
(Spitzer team) \\
M.~Pawlak$^{5}$,
M.~K.~Szyma{\'n}ski$^{5}$,
J.~Skowron$^{5}$,
P.~Mr{\'o}z$^{5}$,
S.~Koz{\l}owski$^{5}$,
{\L}.~Wyrzykowski$^{5}$, 
P.~Pietrukowicz$^{5}$,
G.~Pietrzy{\'n}ski$^{5}$,
I.~Soszy{\'n}ski$^{5}$,
K.~Ulaczyk$^{14}$ \\
(OGLE group) \\
J.-Y.~Choi\altaffilmark{6},
H.~Park\altaffilmark{6},
Y.~K.~Jung\altaffilmark{6},
I.-G.~Shin\altaffilmark{6},
M.~D.~Albrow\altaffilmark{15},
B.-G.~Park\altaffilmark{16},
S.-L.~Kim\altaffilmark{16},
C.-U.~Lee\altaffilmark{16},
S.-M.~Cha\altaffilmark{16,17},
D.-J.~Kim\altaffilmark{16,17},
Y.~Lee\altaffilmark{16,17} \\
(KMTNet group) \\
M.~Friedmann\altaffilmark{8},
S.~Kaspi\altaffilmark{8},
D.~Maoz\altaffilmark{8} \\
(Wise group) \\
M.~Hundertmark$^{10}$,
R.~A.~Street$^{18}$,
Y.~Tsapras$^{19}$,
D.~M.~Bramich$^{20}$,
A.~Cassan$^{9}$,
M.~Dominik$^{21,d}$,
E.~Bachelet$^{18,20}$,
Subo~Dong$^{22}$,
R.~Figuera~Jaimes$^{21,23}$,
K.~Horne$^{21}$,
S.~Mao$^{24}$,
J.~Menzies$^{25}$,
R.~Schmidt$^{19}$,
C.~Snodgrass$^{26}$,
I.~A.~Steele$^{27}$,
J.~Wambsganss$^{19}$ \\
(RoboNet Team) \\
J.~Skottfelt$^{28,10}$,
M.I.~Andersen$^{29}$, 
M.~J.~Burgdorf$^{30}$, 
S.~Ciceri$^{31}$, 
G.~D'Ago$^{4}$, 
D.~F.~Evans$^{32}$, 
S.-H.~Gu$^{33}$, 
T.~C.~Hinse$^{16}$, 
E.~Kerins$^{34}$, 
H.~Korhonen$^{35,10}$, 
M.~Kuffmeier$^{10}$, 
L.~Mancini$^{31}$, 
N.~Peixinho$^{36,37}$, 
A.~Popovas$^{10}$, 
M.~Rabus$^{38}$, 
S.~Rahvar$^{39}$,
R.~Tronsgaard$^{40}$,
G.~Scarpetta$^{3, 11}$, 
J.~Southworth$^{32}$,
J.~Surdej$^{41}$, 
C.~von~Essen$^{40}$,
Y.-B.~Wang$^{33}$, 
O.~Wertz$^{41}$ \\
(MiNDSTEp group) 
}

\affil{
$^{1}$ Department of Astronomy, Ohio State University, 140 W. 18th Ave., Columbus, OH  43210, USA \\
$^{2}$ NASA Exoplanet Science Institute, MS 100-22, California Institute of Technology, Pasadena, CA 91125, USA \\
$^{3}$ Dipartimento di Fisica ``E. R. Caianiello'', Universit\'a di Salerno, Via Giovanni Paolo II, 84084 Fisciano (SA), Italy \\
$^{4}$ Istituto Internazionale per gli Alti Studi Scientifici (IIASS), Via G. Pellegrino 19, 84019 Vietri sul Mare (SA), Italy \\
$^{5}$ Warsaw University Observatory, Al. Ujazdowskie 4, 00-478 Warszawa, Poland \\
$^{6}$ Department of Physics, Chungbuk National University, Cheongju 361-763, Republic of Korea \\
$^{7}$ Jet Propulsion Laboratory, California Institute of Technology, 4800 Oak Grove Drive, Pasadena, CA 91109, USA \\
$^{8}$ School of Physics and Astronomy, Tel-Aviv University, Tel-Aviv 69978, Israel \\
$^{9}$ Sorbonne Universit\'es, UPMC Univ Paris 6 et CNRS, UMR 7095, Institut d'Astrophysique de Paris, 98 bis bd Arago, 75014 Paris, France \\
$^{10}$ Niels Bohr Institute \& Centre for Star and Planet Formation, University of Copenhagen, {\O}ster Voldgade 5, 1350 Copenhagen, Denmark\\
$^{11}$ Istituto Nazionale di Fisica Nucleare, Sezione di Napoli, Italy \\
$^{12}$ Turitea Observatory, Palmerston North, New Zealand \\
$^{13}$ Harvard-Smithsonian Center for Astrophysics, 60 Garden St., Cambridge, MA 02138, USA \\
$^{14}$ Department of Physics, University of Warwick, Gibbet Hill Road, Coventry, CV4 7AL, UK \\
$^{15}$ University of Canterbury, Department of Physics and Astronomy, Private Bag 4800, Christchurch 8020, New Zealand \\
$^{16}$ Korea Astronomy and Space Science Institute, Daejon 305-348, Republic of Korea \\
$^{17}$ School of Space Research, Kyung Hee University, Yongin 446-701, Republic of Korea \\
$^{18}$ Las Cumbres Observatory Global Telescope Network, 6740 Cortona Drive, suite 102, Goleta, CA 93117, USA \\
$^{19}$ Astronomisches Rechen-Institut, Zentrum f{\"u}r Astronomie der Universit{\"a}t Heidelberg (ZAH), 69120 Heidelberg, Germany \\
$^{20}$ Qatar Environment and Energy Research Institute (QEERI), HBKU, Qatar Foundation, Doha, Qatar \\
$^{21}$ SUPA, School of Physics \& Astronomy, University of St Andrews, North Haugh, St Andrews KY16 9SS, UK \\
$^{22}$ Kavli Institute for Astronomy and Astrophysics, Peking University, Yi He Yuan Road 5, Hai Dian District, Beijing 100871, China \\
$^{23}$ European Southern Observatory, Karl-Schwarzschild Stra\ss{}e 2, 85748 Garching bei M\"{u}nchen, Germany,\\
$^{24}$ National Astronomical Observatories, Chinese Academy of Sciences, 100012 Beijing, China \\
$^{25}$ South African Astronomical Observatory, PO Box 9, Observatory 7935, South Africa \\
$^{26}$ Planetary and Space Sciences, Department of Physical Sciences, The Open University, Milton Keynes, MK7 6AA, UK \\
$^{27}$ Astrophysics Research Institute, Liverpool John Moores University, Liverpool CH41 1LD, UK \\
$^{28}$ Centre for Electronic Imaging, Department of Physical Sciences, The Open University, Milton Keynes, MK7 6AA, UK \\
$^{29}$ Niels Bohr Institute, University of Copenhagen, Juliane Mariesvej 30, 2100 Copenhagen {\O}, Denmark \\
$^{30}$ Meteorologisches Institut, Universit{\"a}t Hamburg, Bundesstra\ss{}e 55, 20146 Hamburg, Germany \\
$^{31}$ Max Planck Institute for Astronomy, K{\"o}nigstuhl 17, 69117 Heidelberg, Germany, \\
$^{32}$ Astrophysics Group, Keele University, Staffordshire, ST5 5BG, UK, \\
$^{33}$ Yunnan Observatories, Chinese Academy of Sciences, Kunming 650011, China,\\
$^{34}$ Jodrell Bank Centre for Astrophysics, School of Physics and Astronomy, University of Manchester, Oxford Road, Manchester M13 9PL, UK, \\
$^{35}$ Finnish Centre for Astronomy with ESO (FINCA), V{\"a}is{\"a}l{\"a}ntie 20, FI-21500 Piikki{\"o}, Finland, \\
$^{36}$ Unidad de Astronom{\'{\i}}a, Fac. de Ciencias B{\'a}sicas, Universidad de Antofagasta, Avda. U. de Antofagasta 02800, Antofagasta, Chile, \\
$^{37}$ CITEUC -- Centre for Earth and Space Science Research of the University of Coimbra, Observat\'orio Astron\'omico da Universidade de Coimbra, 3030-004 Coimbra, Portugal \\
$^{38}$ Instituto de Astrof\'isica, Facultad de F\'isica, Pontificia Universidad Cat\'olica de Chile, Av. Vicu\~na Mackenna 4860, 7820436 Macul, Santiago, Chile,\\
$^{39}$ Department of Physics, Sharif University of Technology, PO Box 11155-9161 Tehran, Iran,\\
$^{40}$ Stellar Astrophysics Centre, Department of Physics and Astronomy, Aarhus University, Ny Munkegade 120, 8000 Aarhus C, Denmark, \\
$^{41}$ Institut d'Astrophysique et de G\'eophysique, All\'ee du 6 Ao\^ut 17, Sart Tilman, B\^at. B5c, 4000 Li\`ege, Belgium.
}

\altaffiltext{*}{Email address: weizhu@astronomy.ohio-state.edu}
\altaffiltext{a}{Sagan Visiting Fellow.}
\altaffiltext{b}{NASA Postdoctoral Program Fellow.}
\altaffiltext{c}{Sagan Fellow.}
\altaffiltext{d}{Royal Society University Research Fellow}

\submitted{Submitted to ApJ}

\begin{abstract}
    We report on the mass and distance measurements of two single-lens events from the 2015 \emph{Spitzer} microlensing campaign. With both finite-source effect and microlens parallax measurements, we find that the lens of OGLE-2015-BLG-1268 is very likely a brown dwarf. Assuming that the source star lies behind the same amount of dust as the Bulge red clump, we find the lens is a $45\pm7$ $M_{\rm J}$ brown dwarf at $5.9\pm1.0$ kpc. The lens of of the second event, OGLE-2015-BLG-0763, is a $0.50\pm0.04$ $M_\odot$ star at $6.9\pm1.0$ kpc. We show that the probability to definitively measure the mass of isolated microlenses is dramatically increased once simultaneous ground- and space-based observations are conducted.
\end{abstract}


\section{Introduction} \label{sec:introduction}

Although gravitational microlensing can, in principle, detect faint or even dark objects \citep{Paczynski:1986,Gould:2000}, it is difficult to exclusively determine the mass of the lens object without the measurement of two second-order microlensing parameters: the angular Einstein radius $\theta_{\rm E}$ and the microlens parallax $\pi_{\rm E}$. 
Once $\theta_{\rm E}$ and $\pi_{\rm E}$ are measured, the lens mass $M_{\rm L}$ and the lens-source relative parallax are given by \citep{Gould:1992}
\begin{equation}
M_{\rm L} = \frac{\theta_{\rm E}}{\kappa \pi_{\rm E}}\ ,\quad
\pi_{\rm rel}\equiv \frac{\rm AU}{D_{\rm L}}-\frac{\rm AU}{D_{\rm S}} = \pi_{\rm E} \theta_{\rm E}\ .
\end{equation}
Here
\begin{equation}
\kappa \equiv \frac{4G}{c^2 \rm AU} \approx 8.14 \frac{\rm mas}{M_\odot}\ ,
\end{equation}
and $D_{\rm L}$ and $D_{\rm S}$ are the distances to the lens and the source star, respectively. The main method to measure the Einstein radius $\theta_{\rm E}$ is via the finite-source effect. Such an effect arises because the observed magnification is the integration of the magnification pattern over the face of the source. Therefore, if that magnification pattern has a non-zero second derivative, such as when the source crosses a caustic (where the magnification diverges to infinity), the physical size of the source limits the observed magnification, leading to a rounded feature in the light curve whose width directly reflects the source size \citep{Gould:1994a,WittMao:1994}. The microlens parallax $\pi_{\rm E}$ is measured through observations from either a single accelerating platform \citep{Gould:1992,Honma:1999,Gould:2013} or two well-separated observatories \citep{Refsdal:1966,Gould:1994b,Gould:1997}. With one of these two parameters, one can only obtain a statistical estimate of the lens mass \citep{Yee:2015a,CalchiNovati:2015a}, but this cannot yield unambiguous results for individual specific cases.

\begin{figure*}[!ht]
\epsscale{1.}
\centering
\plotone{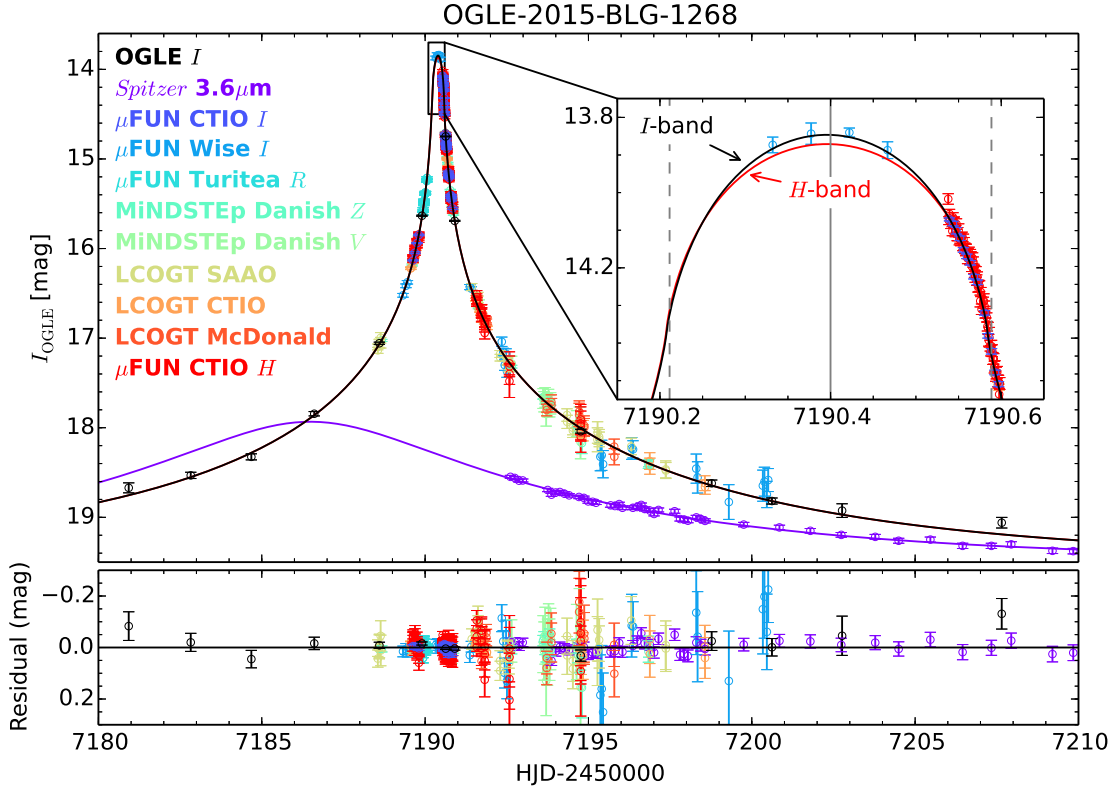}
\caption{Ground-based and \emph{Spitzer} data and best-fit model light curves for OGLE-2015-BLG-1268. The inset shows details around the peak as seen from the ground, with best-fit model seen in the $I$ and $H$ bands shown with different colors. The gray solid line indicates the peak of the event as seen from the ground, and the two dashed vertical lines indicate the two contacts between the point-like lens and the finite-size source. Ground-based data points with uncertainties $>0.2$ mag are suppressed in the figure to avoid clutter, but are included in the modeling.
\label{fig:ob1268-lc}}
\end{figure*}

If only ground-based observations can be obtained, such a method to exclusively determine the lens mass is not very efficient in cases of planet or binary microlensing \citep{MaoPaczynski:1991,GouldLoeb:1992}, and it becomes extremely inefficient for single lenses. In order for the finite source effect to be observed so as to determine $\theta_{\rm E}$, the source should cross or closely pass by the caustic structure of the lens. In the case of a single lens, the caustic is a single point, so if the finite source effect is observed, the apparent alignment between the source and the lens must be so perfect that the lens transits the source. Hence, as in a transit, the feature in the light curve is essentially the face of the source resolved in time, and as a result the points of first and fourth contact can also be seen in the light curve (see, for example, the inset of Figure~\ref{fig:ob1268-lc}). Even in cases in which a finite-source effect is detected, an unambiguous measurement of the lens mass is still hard to achieve, because the parallax signal is intrinsically small and observationally hard to detect \citep{Gould:2009,Yee:2009,GouldYee:2013}.

The emergence of space-based microlensing has seen improvement in the mass measurements of microlensing planets \citep{Udalski:2015a,Street:2015} and binaries \citep{Dong:2007,Zhu:2015a,Shvartzvald:2015}. Here we demonstrate with two examples that such space-based observations also increase the probability to measure the mass of single isolated lenses. Within the 170 microlensing events that were selected for \emph{Spitzer} observations in the 2015 \emph{Spitzer} microlensing program \citep{Gould:2014}, we find two single-lens events that show reliable detection of the finite-source effect. Such an effect is observed only in the ground-based data in OGLE-2015-BLG-1268, and only in the \emph{Spitzer} data in OGLE-2015-BLG-0763. In both cases, the combination of \emph{Spitzer} data and ground-based data sets provides measurement of parallax. Therefore, mass measurements of both microlenses are ensured.

This paper is constructed as follows. The observations of the two events, OGLE-2015-BLG-1268 and OGLE-2015-BLG-0763, are summarized in Section~\ref{sec:observations}; in Section~\ref{sec:modelling} we describe the light curve modeling process; the physical interpretations of both events are given in Section~\ref{sec:physics}; and finally in Section~\ref{sec:discussion} we discuss the interesting findings from these two events, and implications to future space-based microlensing observations.

\section{Observations} \label{sec:observations}

A general description of the \emph{Spitzer} observations and ground-based follow-up strategy for the 2015 \emph{Spitzer} microlensing campaign can be found in \citet{Yee:2015b} and \citet{Street:2015}. In brief, events were selected for \emph{Spitzer} observations, if 1) they showed or were likely to show significant sensitivity to planets, i.e., events with high peak magnifications \citep{GriestSafizadeh:1998} and/or events with intensive survey observations; and 2) could probably yield parallax measurement. Most \emph{Spitzer} targets were assigned the default 1-per-day cadence, except that short or relatively high magnification events were given higher cadence. These observing protocols were submitted on Monday for observations roughly Thursday through Wednesday for each of the six weeks of the campaign. Ground-based follow-up observations were taken mostly on events that were not heavily monitored by surveys, or when surveys could not observe due to weather or Moon passage \citep[e.g., OGLE-2015-BLG-0966,][]{Street:2015}. All ground-based data were reduced using standard algorithms (image subtraction/DoPhot), and the \emph{Spitzer} data were reduced using the new algorithm presented in \citet{CalchiNovati:2015b}.

Below we give detailed descriptions of the observations for these two particular events.

\subsection{OGLE-2015-BLG-1268}
The microlensing event OGLE-2015-BLG-1268 was first alerted by the Optical Gravitational Lens Experiment (OGLE) collaboration through the Early Warning System (EWS) real-time event detection system \citep{Udalski:1994,Udalski:2003} on 2015 June 6, based on observations with the 1.4 deg$^2$ camera on its 1.3m Warsaw Telescope at the Las Campanas Observatory in Chile. With equatorial coordinates (RA, Dec)$_{2000} = (17^{\rm h}56^{\rm m}49\fs77,\ -21\arcdeg53\arcmin57\farcs6$) and Galactic coordinates $(l,b)_{2000}=(7\fdg41,1\fdg42)$, this event lies in the OGLE-IV field BLG642 \citep{Udalski:2015b}, meaning that it is observed by OGLE with a cadence less than once per two nights, and it was not monitored by the MOA or KMTNet surveys.

Because of the very sparse observations from the survey team, follow-up teams including the Microlensing Follow-Up Network \citep[$\mu$FUN,][]{Gould:2010}, Microlensing Network for the Detection of Small Terrestrial Exoplanets \citep[MiNDSTEp,][]{Dominik:2010} and RoboNet \citep{Tsapras:2009} also observed this event fairly intensively during its peak as seen from Earth, in order to support the \emph{Spitzer} program and maximize the sensitivity to potential planets \citep{Yee:2015b,Zhu:2015b}. One remarkable feature about these follow-up observations is that the CTIO SMARTS telescope used by the $\mu$FUN team could simultaneously obtain $H$ band images while $I$ band observations were taken \citep{Depoy:2003}, and these $H$ band observations turned out to be important in characterizing the source star (see next section). Based on the observations taken by 2015 June 16, A.G. issued an alert to the community that this event was deviating from the point-lens point-source model. V.B. later on pointed out that this deviation could be fit by a point-lens finite-source model. 

Event OGLE-2015-BLG-1268 was selected for \emph{Spitzer} observations on 2015 June 14, i.e., $\sim2$ days before it reached its peak magnification as seen from Earth. It was classified as a ``subjective'' event, because it by then did not (in fact, never could) meet the objective criteria set by \citet{Yee:2015b}. It was selected primarily because it was going to peak with relatively high magnification in the coming week. With a predicted peak magnification $A_{\rm max,\oplus}>6$ (1-$\sigma$ lower limit) by the time of selection, this event qualified for bonus observations \citep{Yee:2015b}, and so was assigned 4-per-day cadence for its first week of observations. Starting from its second week, the \emph{Spitzer} cadence went down to the default value (1-per-day), and this cadence continued for two weeks until it met the criteria set by \citet{Yee:2015b} for stopping observations. In total, \emph{Spitzer} observed this event 48 times, each with 6 dithered 30s exposures.

The data for this event are shown in Figure~\ref{fig:ob1268-lc}. All data sets have been aligned to the OGLE-IV $I$ magnitude according to the best-fit model given in Section~\ref{sec:modelling-1268}.

\begin{deluxetable}{lc}
\centering
\tablecaption{Best-fit parameters of OGLE-2015-BLG-1268.}
\tablewidth{0pt}
\tablehead{Parameters & Values}
\startdata
$\chi^2$/dof & 700.7/693 \\
$t_0$ (HJD$^\prime-7190$) & 0.3992 $\pm$ 0.0016 \\
$u_0$ ($10^{-4}$) & 0.5 $\pm$ 3.2 \\
$t_{\rm E}$ (days) & 17.5 $\pm$ 0.7 \\
$\rho$ ($10^{-2}$) & 1.08 $\pm$ 0.04 \\
$\pi_{\rm E,N}$ & $\pm0.24$ $\pm$ 0.07 \\
$\pi_{\rm E,E}$ & 0.25 $\pm$ 0.07 \\
Blend$^a$ & 0.029 $\pm$ 0.039 \\

\enddata
\tablecomments{$^a$ Blending fraction in the OGLE-IV $I$ band.}
\label{tab:ob1268}
\end{deluxetable}

\subsection{OGLE-2015-BLG-0763}
This event was first alerted by the OGLE collaboration on 2015 April 22. With equatorial coordinates (RA, Dec)$_{2000} = (17^{\rm h}32^{\rm m}23\fs41,\ -29\arcdeg18\arcmin09\farcs7$) and Galactic coordinates $(l,b)_{2000}=(-1\fdg78,2\fdg25)$, this event lies in the OGLE-IV field BLG613 \citep{Udalski:2015b} and is observed by OGLE with once-per-night cadence. It also lies within the region where the new microlensing survey, Korean Microlensing Telescope Network \citep[KMTNet,][]{Kim:2016} obtained 1-2 observations per day from each of their three 1.6m telescopes as support to the \emph{Spitzer} campaign. Thus, it is also labeled by the KMTNet team as KMT-2015-BLG-0187.

\begin{figure*}[!ht]
\epsscale{1.}
\centering
\plotone{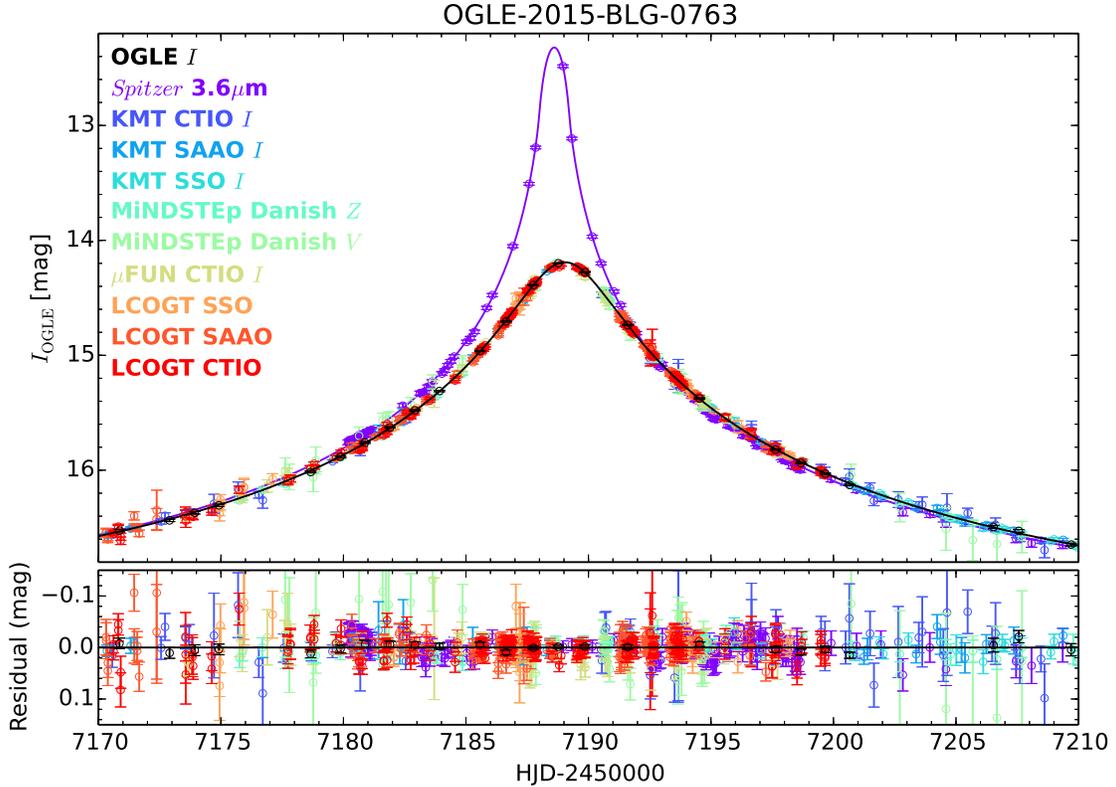}
\caption{Ground-based and \emph{Spitzer} data and best-fit model light curves for OGLE-2015-BLG-0763. The finite source effect is only seen in the \emph{Spitzer} data. Ground-based data points with uncertainties $>0.2$ mag are suppressed in the figure to avoid clutter, but are included in the modeling.
\label{fig:ob0763-lc}}
\end{figure*}

Events such as this, with OGLE-IV cadence of 1-per-day and lying in non-prime KMTNet fields, are treated by \citet{Yee:2015b} as having low survey coverage in all respects. That is, they are not eligible for objective selection, and once they are selected subjectively, follow-up observations are strongly encouraged. This event hence received a fairly large amount of observations from $\mu$FUN, RoboNet, and MiNDSTEp, besides the survey observations, although none of these ground-based data showed deviation from the standard point source model.

Event OGLE-2015-BLG-0763 was selected by the \emph{Spitzer} team on 2015 May 19, when the Bulge could not yet be seen from \emph{Spitzer} due to its Sun-angle limit. Nevertheless, at that time it was selected as a future \emph{Spitzer} target in order to achieve high sensitivity to planets. As specified in \citet{Yee:2015b} and enacted in \citet{Zhu:2015b}, when measuring planet sensitivity or detecting planets, only the portion of the light curve taken after the subjective selection of an event may be considered 
\footnote{Unless it is later found to meet objective selection criteria.},
in order to isolate any knowledge of the existence of potential planet from the decision making; this is necessary to maintain the objectivity of the sample.
The Bulge became accessible to \emph{Spitzer} starting from June 6 (HJD$'\equiv$ HJD-2450000=7180). Although this event was only assigned 1-per-day cadence, its position, relatively far to the west, made it one of the few targets that could be observed by \emph{Spitzer} in the beginning of the Bulge window, and thus it received 8-per-day cadence in the first week. The cadence dropped to 2-per-day as more targets became accessible to \emph{Spitzer} in the second week. Beginning the third week all targets became accessible to \emph{Spitzer}, and we devoted the extra time to relatively high magnification events \citep[see][for detailed description]{Street:2015}. Then the cadence for OGLE-2015-BLG-0763 was increased to 8-per-day this week due to its high peak magnification as seen from Earth. The cadence went back to the default one (1-per-day) in the following weeks until it moved out of \emph{Spitzer} window on HJD$'=7218$. In total, 142 \emph{Spitzer} observations were obtained.

The \emph{Spitzer} data for the six epochs nearest the peak of OGLE-2015-BLG-0763 are affected by saturation and/or non-linearity. This is quite apparent from the online data presented by \citet{CalchiNovati:2015b}.  For this event, we have therefore re-reduced the data by taking into account the specific characteristics of channel 1 (at 3.6 $\mu$m) of the IRAC camera.  We treat all pixels with counts greater than 70\% of the allowed maximum as ``corrupted'', i.e., as saturated and/or non-linear.  In addition, all pixels that are interior to corrupted pixels are also treated as corrupted, since the saturation of these pixels can falsely lead to ``normal'' flux counts. This leads to masking seven to nine pixels for each of these six dithered images of the epoch closest to peak and to masking up to seven pixels for each of the six dithered images for the remaining 5 affected epochs.  The remaining pixels (i.e., those below 70\% of full well and not interior to corrupted pixels) are treated as normal, since they are essentially unaffected by the non-linearities in neighboring pixels.  After the masking of the inner saturated pixels, the fit is therefore carried out in standard fashion.  We note that our 70\% threshold is conservative with respect to the 90\% value in IRAC documentation, and was determined by us empirically from inspection of the raw data and by studying the stability of the photometry PRF fitting solution.
\footnote{We have tested that the results remain the same even though higher threshold values (up to 90\%) are used.}

The data for this event are shown in Figure~\ref{fig:ob0763-lc}. Again, all data sets have been aligned to the OGLE-IV $I$ magnitude according to the best-fit model given in Section~\ref{sec:modelling-0763}.

\begin{deluxetable}{lcc}
\centering
\tablecaption{Best-fit parameters of OGLE-2015-BLG-0763. The other two possible solutions, $(+,+)$ and $(-,-)$, are disfavored by $\Delta \chi^2=15$, and have $\rho \le 0.01$ (2-$\sigma$ limit) that is inconsistent with the blend constraint (see Section~\ref{sec:physics}).}
\tablewidth{0pt}
\tablehead{Parameters & $(+,-)$ & $(-,+)$}
\startdata
$\chi^2$/dof & 1469.8/1463 & 1468.7/1463 \\
$t_0$ (HJD$^\prime-7189$) & 0.038 $\pm$ 0.005 & 0.045 $\pm$ 0.005 \\
$u_0$ ($10^{-2}$) & 5.92 $\pm$ 0.06 & -5.99 $\pm$ 0.06 \\
$t_{\rm E}$ (days) & 33.09 $\pm$ 0.24 & 32.78 $\pm$ 0.25 \\
$\rho$ ($10^{-2}$) & 2.18 $\pm$ 0.05 & 2.17 $\pm$ 0.05 \\
$\pi_{\rm E,N}$ & -0.0684 $\pm$ 0.0007 & 0.0707 $\pm$ 0.0008 \\
$\pi_{\rm E,E}$ & 0.0174 $\pm$ 0.0002 & 0.0083 $\pm$ 0.0002 \\
Blend$^a$ & -0.035 $\pm$ 0.010 & -0.049 $\pm$ 0.012 \\

\enddata
\tablecomments{$^a$ Blending fraction in the OGLE-IV $I$ band.}
\label{tab:ob0763}
\end{deluxetable}

\section{Light Curve Analyses} \label{sec:modelling}

\subsection{Modeling Process}
For each event, we perform a Markov Chain Monte Carlo (MCMC) analysis to find the best-fit parameters and the associated probability density functions using the \texttt{emcee} ensemble sampler \citep{ForemanMackey:2013}. The formalism to incorporate the finite-source effect in the single-lens case is given in \citet{Yoo:2004}. For each microlensing event, there will be six event parameters: time of maximum magnification as seen from Earth $t_0$, the impact parameter in the absence of parallax as seen from Earth $u_0$, the event timescale $t_{\rm E}$, the source radius scaled to the angular Einstein radius $\rho$, and the two components of microlens parallax along the north and east direction $\pi_{\rm E,N}$ and $\pi_{\rm E,E}$. For each data set on that event, there are two flux parameters ($f_s$, $f_b$). In addition, we have two limb-darkening coefficients ($\tilde{q}_{1,\lambda}$, $\tilde{q}_{2,\lambda}$) described in the next section. If these are also set free, there are two additional parameters for each bandpass.

For each event that is investigated below, we report the six event parameters as well as the blending fraction in the OGLE-IV $I$ band. This blending fraction, defined as $f_b/(f_s+f_b)$, can be used to distinguish between degenerate models, as we will see below.

\subsection{Limb Darkening Effect}
For the present cases, especially OGLE-2015-BLG-1268, many of our observations were taken when the alignment between the source and the lens was closer than roughly the scaled source size $\rho$, so the generally adopted linear limb-darkening law is not adequate. Therefore, we choose the two-parameter square root limb darkening law
\begin{equation} \label{eq:sqrtld}
S_\lambda (\mu) = \bar{S}_\lambda \left[ 1-\Gamma_\lambda \left(1-\frac{3}{2}\mu\right) - \Lambda_\lambda \left(1-\frac{5}{4}\sqrt{\mu}\right) \right]\ ,
\end{equation}
in which $\mu$ is the cosine of the angle between the line of sight and the emergent intensity, and $\Gamma_\lambda$ and $\Lambda_\lambda$ are the limb-darkening coefficients at wavelength $\lambda$. Because of our requirement that the total flux, $F_{\rm tot,\lambda}=\pi \theta_\star^2 \bar{S}_\lambda$, should be conserved for a given source angular size $\theta_\star$, Equation~(\ref{eq:sqrtld}) differs from the standard square root limb-darkening law described by $(c_\lambda,d_\lambda)$ in \citet{DiazCordoves:1992}. The transformations between limb-darkening coefficients ($\Gamma_\lambda,\Lambda_\lambda$) and ($c_\lambda,d_\lambda$) can be found in \citet{Fields:2003}. 

In cases in which very little knowledge of the source star can be obtained otherwise, one may want to fit for ($\Gamma_\lambda,\Lambda_\lambda$). However, simply leaving ($\Gamma_\lambda,\Lambda_\lambda$) free will very likely lead to unphysical stellar brightness profiles. Therefore, we reparameterize the limb-darkening coefficients ($\Gamma_\lambda,\Lambda_\lambda$) using the method proposed by \citet{Kipping:2013},
\begin{equation} \label{eq:q1}
\tilde{q}_{1,\lambda} \equiv \left(\Gamma_\lambda + \Lambda_\lambda\right)^2\ ,
\end{equation}
\begin{equation} \label{eq:q2}
\tilde{q}_{2,\lambda} \equiv \frac{7\Lambda_\lambda}{12(\Gamma_\lambda+\Lambda_\lambda)}\ .
\end{equation}
The inverse transformation is given by
\begin{equation}
\Gamma_\lambda = \sqrt{\tilde{q}_{1,\lambda}} \left(1-\frac{12}{7}\tilde{q}_{2,\lambda}\right)\ ,
\end{equation}
\begin{equation}
\Lambda_\lambda = \frac{12}{7} \sqrt{\tilde{q}_{1,\lambda}}\tilde{q}_{2,\lambda}\ .
\end{equation}
As \citet{Kipping:2013} has shown, uniform samplings in the interval [0, 1] for both $\tilde{q}_{1,\lambda}$ and $\tilde{q}_{2,\lambda}$ can ensure physically meaningful stellar intensity profiles 
\footnote{Positive intensity everywhere, and monotonically decreasing intensity from the center to the limb.}
as well as being efficient.
\footnote{Note that ($\tilde{q}_{1,\lambda}$, $\tilde{q}_{2,\lambda}$) given by Equations~(\ref{eq:q1}) and (\ref{eq:q2}) is not the only parameterization to implement the idea proposed by \citet{Kipping:2013}.}

\subsection{OGLE-2015-BLG-1268} \label{sec:modelling-1268}

In the case of OGLE-2015-BLG-1268, because we were unable to fully understand the source star with only photometric data (see Section~\ref{sec:physics-1268}), we also fit for the limb-darkening coefficients. Since only the $I$ and $H$ band data were relevant when the finite-source effect is prominent, we only set the limb-darkening coefficients of these two bands free. Furthermore, since the \emph{Spitzer} data only captured the falling tail of the light curve, a constraint on the source flux in the \emph{Spitzer} 3.6 $\mu$m band ($L_{\rm spitzer}$) is necessary in order to better measure the parallax \citep{CalchiNovati:2015a,CalchiNovati:2015b}. This is done by conducting a color-color regression between $I-H$ and $I-L_{\rm spitzer}$ using stars with color and magnitude similar to the clump.

\begin{figure*}
\centering
\plotone{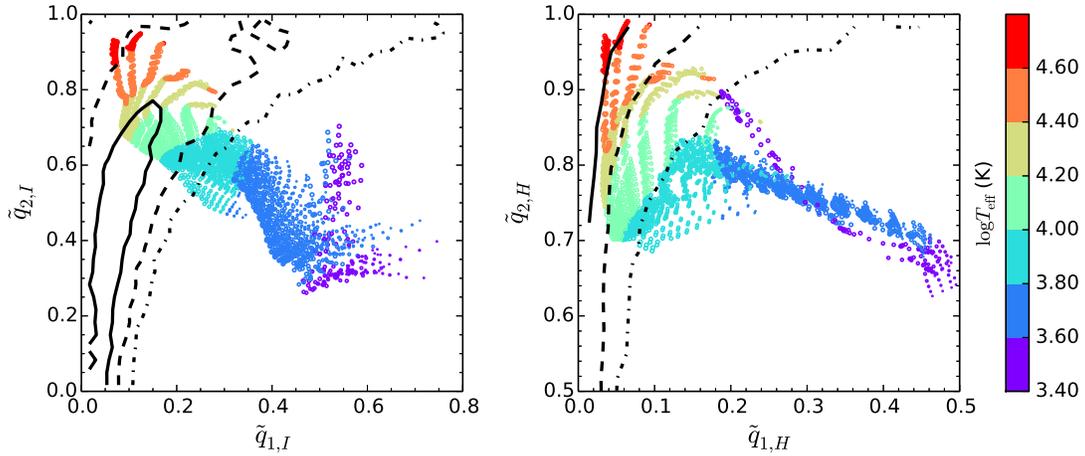}
\caption{The $\Delta \chi^2=1$, 4 and 9 constraints (solid, dashed and dot-dashed curves, respectively) on the limb-darkening coefficients in the $I$ band (left) and $H$ band (right) in the case of OGLE-2015-BLG-1268. Open circles are theoretical values taken from \citet{Claret:2000} after the transformation from $(c,d)$ to $(\tilde{q}_1,\tilde{q}_2)$, and they are color coded with the stellar effective temperature. The constraints on limb-darkening coefficients effectively rule out a source with $T_{\rm eff}<8000$ K.
\label{fig:ob1268-ldcs}}
\end{figure*}

Our best-fit model to this event is shown in Figure~\ref{fig:ob1268-lc}, and best-fit parameters with 1-$\sigma$ intervals are given in Table~\ref{tab:ob1268}. Note that in Figure~\ref{fig:ob1268-lc}, the $H$ band model shows deviations from the $I$ band model, which are obvious on the top and still noticeable at the limb. Characterizing events via such deviations for point-lens \citep{GouldWelch:1996} and planetary \citep{GaudiGould:1997} events was the original motivation for building the dichroic ANDICAM camera \citep{Depoy:2003}.
We show in Figure~\ref{fig:ob1268-ldcs} the $\Delta \chi^2=1$, 4 and 9 contours between $(\tilde{q}_{1,\lambda},\ \tilde{q}_{2,\lambda})$ on the top of the theoretical values from \citet{Claret:2000}. The constraint on the $I$ band limb-darkening coefficients $(\tilde{q}_{1,I},\ \tilde{q}_{2,I})$ favors hot stars ($T_{\rm eff}\ge8000$ K) at 2-$\sigma$ level, and this is consistent with the less constrained $H$ band limb-darkening coefficients.

Since the impact parameter as seen from Earth is extremely small (indeed, consistent with zero) compared to the that seen from \emph{Spitzer}, the generic four-fold degeneracy in single-lens events collapses to two-fold, i.e., to a degeneracy only in the direction of $\bdv{\pi}_{\rm E}$ (and not its magnitude). However, this degeneracy in direction has no effect on the mass and distance measurement \citep{GouldYee:2012}.

\subsection{OGLE-2015-BLG-0763} \label{sec:modelling-0763}
For OGLE-2015-BLG-0763, we adopt the limb-darkening coefficients that correspond to the well characterized source (see Section~\ref{sec:physics-0763}), so that only event parameters and flux parameters are free in the modeling. The best-fit model is shown in Figure~\ref{fig:ob0763-lc}, and the best-fit parameters together with 1-$\sigma$ intervals for the two adopted solutions, $(+,-)$ and $(-,+)$, are given in Table~\ref{tab:ob0763}. Parameters of the other possible solutions, i.e., $(+,+)$ and $(-,-)$ solutions, are similar except that only an upper limit on the source size, $\rho\le0.01$ at $\Delta \chi^2=4$ or 2-$\sigma$ level, can be obtained, meaning that the finite-source effect is not securely detected. Nevertheless these two solutions are rejected, partly because they are disfavored by $\Delta \chi^2=15$, and also because the derived physical parameters result in inconsistency with observations (see Section~\ref{sec:physics-0763}).

The two adopted solutions have different parallax vector $\bdv{\pi}_{\rm E}$ but similar amplitude $\pi_{\rm E}$, as is shown in Table~\ref{tab:ob0763}. As in the previous case, for the purpose of mass and distance determination this means that the four-fold degeneracy is also broken here. We show in Section~\ref{sec:discussion} that this is not coincidental, but is rather almost inevitable for mass measurements of single-lens events with space-based observations.

\begin{figure*}
\epsscale{1.}
\centering
\plottwo{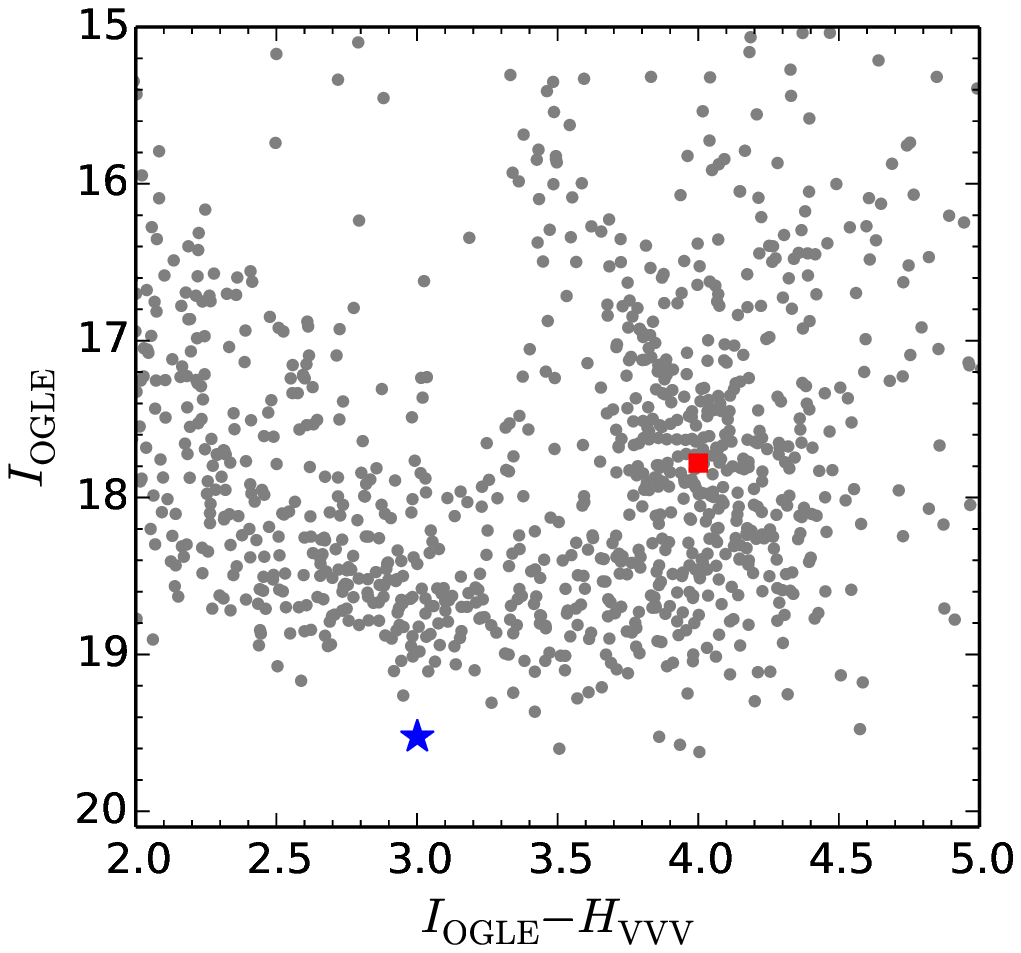}{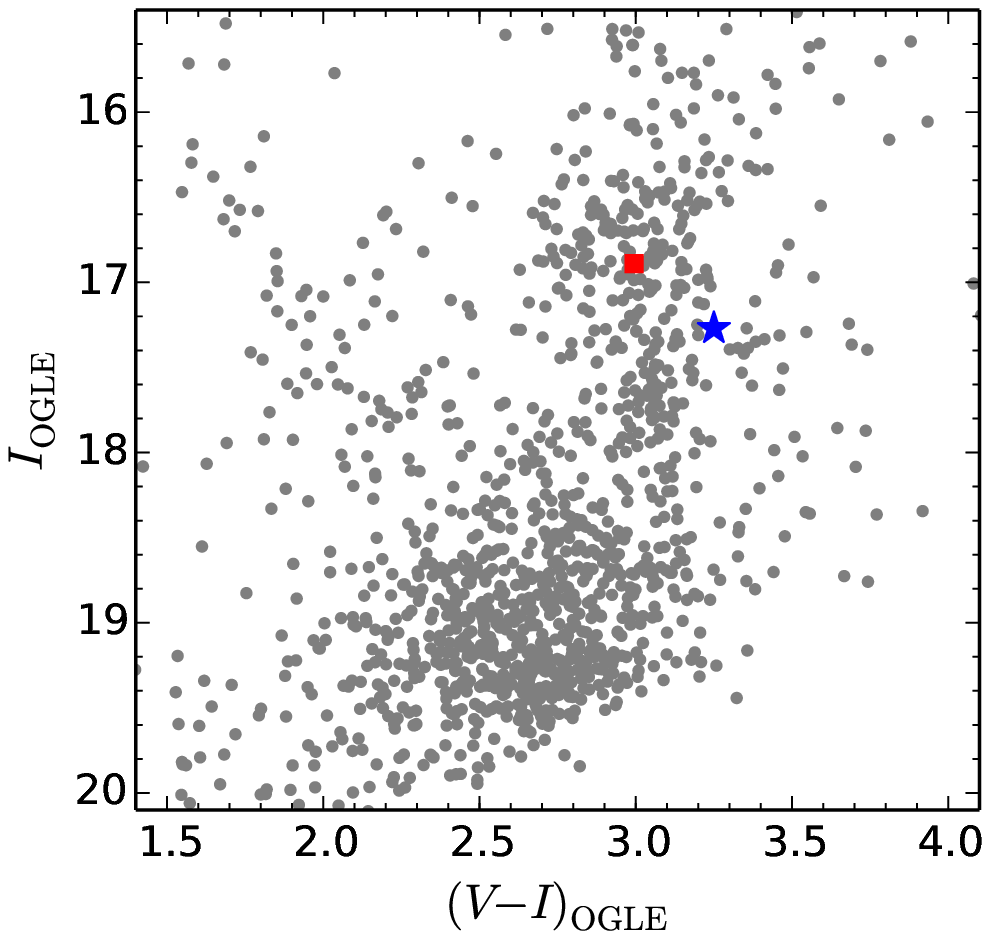}
\caption{The color-magnitude diagrams (CMD) used to characterize the source of OGLE-2015-BLG-1268 (left) and OGLE-2015-BLG-0763 (right). In both cases, the centroid of the red clump is indicated by a filled red square, and the position of the source is marked by a blue asterisk.
\label{fig:cmds}}
\end{figure*}

\section{Physical Interpretations} \label{sec:physics}

\subsection{OGLE-2015-BLG-1268L: A Brown Dwarf in the Inner Disk} \label{sec:physics-1268}

The extinction in the region where this event lies is so high that the routine OGLE-IV $V$ band images are not deep enough to resolve the red giant clump in the Bulge. In order to determine the color and magnitude of the source star following the standard routine \citep{Yoo:2004}, we then construct an $I-H$ vs. $I$ color-magnitude diagram (CMD) in the following way: we obtain OGLE-IV $I$ band magnitudes for stars within $2'$ separation from the event, and cross-match to the VISTA Variables in the V\'ia L\'actea Survey \citep[VVV,][]{Saito:2012} to obtain their $H$ band magnitudes. 
The position of the source star on this CMD is found to be
\begin{equation}
(I-H,\ I)_s = (2.997\pm0.003,\ 19.54\pm0.04)\ ,
\end{equation}
in which $I_s$ comes directly from the modeling and $H_s$ is determined by transforming the CTIO instrumental $H$-band source flux that is derived from the model to the standard VVV system using comparison stars. The centroid of the red clump on this CMD is at
\begin{equation}
(I-H,\ I)_{\rm cl} = (4.00\pm0.02,\ 17.78\pm0.05)\ .
\end{equation}
The CMD for this event with the positions of the source and the clump centroid indicated is shown on the left panel of Figure~\ref{fig:cmds}.
Comparing to the dereddened color and magnitude of the clump at this particular direction, $(I-H,\ I)_{\rm cl,0}=(1.29,\ 14.23)$ \citep{Bensby:2013,Nataf:2013}, we determine the extinction and reddening to be $A_I=3.6$ and $E(I-H)=2.7$, respectively. If all of this extinction applies to the source star, the dereddened source will have 
\begin{equation}
(I-H,\ I)_{\rm s,0}=(0.29\pm0.02,15.99\pm0.08)\ .
\end{equation}
and $M_I=1.65$. These values suggest the source star is a late A/early F dwarf \citep{BessellBrett:1988}. Such stars are rare in the Bulge \citep{Bensby:2013,Poleski:2014}. In principle, one could not exclude the possibility that the source might not lie behind the same amount of dust as the clump does, so that an intrinsically redder and therefore dimmer but closer source star is also allowed by this CMD analysis. However, the limb-darkening coefficients from our light curve modeling (see Figure~\ref{fig:ob1268-ldcs}) also suggest that the source star has $T_{\rm eff}\ge8000$ K when compared to the theoretical values given by \citet{Claret:2000}. Therefore, we conclude that the source star is an A-type dwarf in the Bulge (or at least at a roughly similar distance as the Bulge).

We then use the dereddened source color and magnitude to derive the angular radius of the source star. \citet{Boyajian:2014} provide the following relation to estimate the stellar angular diameter based on the dereddened $I-H$ color and $I$ magnitude
\begin{equation} \label{eq:thetas}
\log{d_{\star}} = a_0 + a_1 (I-H) - 0.2I\ ,
\end{equation}
with $a_0=0.53026\pm0.00077$, $a_1=0.36595\pm0.00079$. Adopting the above dereddened color and magnitude of the source, we find that the source angular radius is
\begin{equation}
\theta_\star = (1.37\pm0.10)\ \mu{\rm as}\ .
\end{equation}
The uncertainty here is dominated by the $7.4\%$ scatter in the color-surface brightness relation (Equation~\ref{eq:thetas}).

When combined with the source radius $\rho=0.0108\pm0.0005$, it yields $\theta_{\rm E}=0.127\pm0.009$ mas. Given the microlens parallax measurement $\pi_{\rm E}=0.35\pm0.04$
\footnote{Note that the uncertainty on $\pi_{\rm E}$ is smaller than the uncertainty on either $\pi_{\rm E,N}$ or $\pi_{\rm E,E}$ (see Table~\ref{tab:ob1268}) because of the anti-correlation between $\pi_{\rm E,N}$ and $\pi_{\rm E,E}$. This anti-correlation emerges due to the constraint on the source flux in $L_{\rm spitzer}$ band. In particular, \citet{GouldYee:2012} showed that for high-magnification events (as seen from Earth), $\pi_{\rm E}$ can be determined from a single space-based observation taken at the ground-based peak (plus one additional late-time observation), provided the space-based source flux is known. In this case, there would be perfect knowledge of the magnitude of $\bdv{\pi}_{\rm E}$ and perfect ignorance of its direction. While the observations of OGLE-2015-BLG-1268 do not satisfy this condition (see Figure~\ref{fig:ob1268-lc}), essentially the same logic applies.}
We then find the lens mass and the lens-source relative parallax
\begin{equation}
M_{\rm L} = (0.045\pm0.007)\ M_\odot;
\quad \pi_{\rm rel} = (0.044 \pm 0.006)\ {\rm mas}\ ,
\end{equation}
and then (adopting $D_{\rm S}=8\pm1$ kpc), a lens distance
\begin{equation}
D_{\rm L} = (5.9\pm1.0)\ {\rm kpc}\ .
\end{equation}
That is, a brown dwarf (BD) in the inner Galactic disk. 

The lens-source relative proper motion, $\mu_{\rm rel}=2.65\pm0.20~\masyr$, is more typical of Bulge lenses ($\mu_{\rm rel}\sim4~\masyr$) than disk lenses ($\mu_{\rm rel}\sim7~\masyr$). However, this value is certainly not inconsistent with a disk lens for two reasons. First, the higher proper motions typical of disk lenses come from the fact that the Sun and the disk lens partake of the same relatively flat Galactic rotation curve, while Bulge sources are on average not moving in the frame of the Galaxy. Hence, disk lenses tend to have similar proper motions as SgrA*. However, in the present case, the source may also be in the disk as indicated by the fact that it is likely to be an A star. If so, it would partake of the same flat rotation curve, which then naturally leads to a low proper motion. Second, one of the two solutions (with $\pi_{\rm E,N}>0$) is approximately in the direction of Galactic rotation, so that the Bulge source would have to be moving at just $\mu_{\rm S}\sim 4~\masyr$ in the direction of Galactic rotation (i.e., $1.3~\sigma$) to account for the observed relative proper motion.

The Galactic coordinates of the event, $(l,b)=(7\fdg41,1\fdg42)$, provide a further indication that the source may be in the disk. That is, even if the source were at roughly the Galactocentric distance, it would lie just $\sim200$~pc from the plane, which is significantly populated by late A/early F stars. Moreover, at this longitude, disk stars account for a significantly larger fraction of stars at these distances than is the case for typical microlensing fields near $l\sim0$. Thus, this possibility must be taken seriously.

Now, if the source were in the disk, this would not in any way affect either $M_{\rm L}$ or $\pi_{\rm rel}$, provided that the source lies behind the same dust column as the clump. This is because the estimate of $\theta_\star$ (and so $\theta_{\rm E}$) depends only on the dereddened color and magnitude of the source and not its distance. However, given the various evidences favoring a disk source, as well as the low Galactic latitude of the event, we must consider the possibility that the source does in fact lie in front of at least some of the dust.

Hence, we develop a more robust way to estimate the lens mass that does not require the source lying behind all the dust. According to the color-surface brightness relation (Equation~\ref{eq:thetas}), one can write the angular source radius $\theta_\star$ (and further $M_{\rm L}$) in terms of the dereddened source color $(I-H)_{\rm s,0}$ and magnitude $I_{\rm s,0}$. If the reddening law derived from the red clump, $R_{IH}=A_I/E(I-H)=1.33$, also applies to the source star, we can express $(I-H)_{\rm s,0}$ and $I_{\rm s,0}$ in terms of $(I-H)_{\rm s}$, $I_{\rm s}$, and a single parameter $\eta$,
\begin{equation} \label{eq:eta}
I_{\rm s,0} = I_{\rm s,0,naive} + \eta A_I\ ,\quad
(I-H)_{\rm s,0} = (I-H)_{\rm s,0,naive} + \eta E(I-H)\ .
\end{equation}
Here $\eta$ is the fraction of total extinction of dust lying behind the source (and in front of the clump), while $I_{\rm s,0,naive}$ and $(I-H)_{\rm s,0,naive}$ are the values we derived above under the naive assumption that $\eta=0$. Combining Equations~(\ref{eq:thetas}) and (\ref{eq:eta}) with the observed $R_{IH}=1.33$ and $E(I-H)=2.7$, one finds that as one varies $\eta$,~$\theta_\star$ changes by
\begin{equation} \label{eq:dthetas}
\Delta \log\theta_\star = \eta E(I-H) (a_1-0.2 R_{IH}) = 0.27\eta\ .
\end{equation}
Thus in particular, our conclusion that the best estimate for the lens mass is in the brown dwarf regime ($M<0.08M_\odot$) rests on the assumption that $\eta<\log(0.08/0.045)/0.27=0.86$. This is very likely simply because nearby sources are extremely rare in microlensing due to low optical depth. Nevertheless, it would be of interest to definitively resolve this issue. This could be done simply by identifying the stellar type of the source via a low-resolution spectrum. This would remove the uncertainty due to the location of the source with respect to the dust. It would also permit one to determine $\theta_\star$ without appealing to the detailed modeling of limb-darkening coefficients from the light curve (Section~\ref{sec:modelling-1268}).

\subsection{OGLE-2015-BLG-0763L: An M-type Main-Sequence Star in the Inner Disk} \label{sec:physics-0763}
We first characterize the source of OGLE-2015-BLG-0763 using the method of \citet{Yoo:2004}. Combining the light curve modeling and a linear regression of OGLE-IV $V$ band on $I$ band flux during the event yield
$(V-I,\ I)_{\rm s} = (3.25,\ 17.27)$. The clump centroid in this $V-I$ vs. $I$ CMD (see the right panel of Figure~\ref{fig:cmds}) is found at $(V-I,\ I)_{\rm cl} = (3.00,\ 16.89)$. After the correction of OGLE-IV non-standard $V$ band \citep{Zhu:2015a}, and combining with the dereddened color \citep{Bensby:2013} and magnitude \citep{Nataf:2013} of the clump at this direction, we derive the color and magnitude of the dereddened source to be $(V-I,\ I)_{\rm s,0} = (1.29,\ 14.78)$. Thus, the source is a giant star in the Bulge. We then convert from $V-I$ to $V-K$ using the empirical color-color relations of \citet{BessellBrett:1988}, apply the color-surface brightness relation of \citet{Kervella:2004}, and find the source angular radius
\begin{equation}
\theta_\star = 6.3\pm0.4\ \mu{\rm as}\ .
\end{equation}
The uncertainty here has taken into account the uncertainty in centroiding the clump $I$-band magnitude in the CMD ($\sim0.05$ mag) and the derivation of the intrinsic source color \citep[0.05 mag,][]{Bensby:2013}.

Given the source size $\rho=0.0218\pm0.0007$ from light curve modeling,
\footnote{Here we combine the results of $(+,-)$ and $(-,+)$ solutions, and the uncertainty takes into account the difference between both solutions and the uncertainty of each solution.}
this indicates an angular Einstein radius $\theta_{\rm E}=0.288\pm0.020$ mas. We then combine this with the microlens parallax parameter $\pi_{\rm E} = 0.0709\pm0.0010$ 
and find the lens mass and distance
\begin{equation}
M_{\rm L} = 0.50\pm0.04\ M_\odot\ ,\quad D_{\rm L} = 6.9\pm1.0\ {\rm kpc}\ .
\end{equation}
In deriving $D_{\rm L}$ we have assumed a Bulge source ($D_{\rm S}=8\pm1$ kpc), which is very likely the case given the location of this event. Given the timescale $t_{\rm E}=32.9\pm0.3$ days, the lens-source relative proper motion is $\mu_{\rm rel}=3.20\pm0.25$ mas/yr, and it is almost due the north/south direction ($\pi_{\rm E,E}\ll |\pi_{\rm E,N}|$).

As is mentioned in Section~\ref{sec:modelling-0763}, the other two solutions $(+,+)$ and $(-,-)$ are disfavored by $\Delta\chi^2=15$ and only provide upper limit of $0.01$ on $\rho$. Regardless of the $\chi^2$ argument, if this set of solutions were correct, the lens would have mass $1.0\ M_\odot$ and distance 6 kpc. First of all, more massive stars are intrinsically rarer. Furthermore, a 1 $M_\odot$ main-sequence star would be $12\%$ as bright as the source star, and therefore contribute 11\% of the total baseline flux. This contradicts the fact that slightly negative blending is detected in this event.
\footnote{The adopted solution, a 0.5 $M_\odot$ star at 6.9 kpc, cannot produce a negative blending either, but this solution is much more plausible given the fact that various effects can lead to slightly negative blending \citep{Smith:2007}.}
Therefore, these two solutions are discarded.

\section{Discussion} \label{sec:discussion}

The mass determination of isolated microlenses requires measurements of at least two of the three important microlensing parameters: the angular Einstein radius $\theta_{\rm E}$, the microlens parallax $\pi_{\rm E}$, and the lens flux. Among them, measuring the lens flux requires the lens to be luminous, so that it will not work for faint or dark lenses, although it may see broad applications in future space-based microlensing experiments \citep{Yee:2015c}. Therefore, measuring $\theta_{\rm E}$ and $\pi_{\rm E}$ together is the only method that works for all types of objects.

In this work, we report on the mass and distance measurements of two single-lens events from our 2015 \emph{Spitzer} microlensing program. The finite-source effect is detected in the ground-based data in OGLE-2015-BLG-1268, and in the \emph{Spitzer} data in OGLE-2015-BLG-0763, ensuring measurement of the angular Einstein radius $\theta_{\rm E}$. The microlens parallax parameter $\pi_{\rm E}$ is measured from a combined fit to the ground- and space-based data. We find that the lens of OGLE-2015-BLG-1268 is very likely a brown dwarf, and probably a 45 $M_{\rm J}$ brown dwarf at 5.9 kpc if the source star is inside the galactic Bulge. The lens of OGLE-2015-BLG-0763 is a 0.5 $M_\odot$ star at 6.9 kpc.

The result that at least two 
\footnote{The single-lens event OGLE-2015-BLG-1482 also shows finite-source effect as seen from \emph{Spitzer} (Chung et al., in prep).}
single-lens events out of the 170 monitored events in the 2015 \emph{Spitzer} program yield mass measurements was unexpected, but turns out to be reasonable.
To measure the mass of isolated microlenses by combining $\theta_{\rm E}$ and $t_{\rm E}$, one needs to detect the finite-source effect and measure the microlens parallax. The first is approximately a geometric probability for the lens to ``transit'' the source star, which is of order $\rho$
\footnote{The observational bias tends to enhance the probability of detecting such finite-source effects in single-lens events that are caused by faint dwarf source stars. See \citet{ZhuGould:2016} for more discussions.}    
, and the second, in the case of two-observatory parallax method \citep{Refsdal:1966,Gould:1994b,Gould:1997}, is limited mostly by the projected separation between the two observatories $D_\perp$. In order to measure the parallax, the two observatories should both be able to detect the same event, but should each see (slightly) different light curves. In the presence of the finite-source effect, this provides the following constraints on the projected Einstein radius $\tilde{r}_{\rm E}\equiv{\rm AU}/\pi_{\rm E}$,
\begin{equation}
    D_\perp \lesssim \tilde{r}_{\rm E} \lesssim \frac{50 D_\perp}{\rho}\ .
\end{equation}
where in the second inequality we have assumed typical photometric uncertainty ($2\%$) during peak \citep{GouldYee:2013}. For a lens with mass $M_{\rm L}$, the above inequalities then provide upper and lower limit on the lens distance for which microlens parallax is detectable (assuming Bulge sources)
\begin{equation} \label{eqn:pirellimits}
    \frac{\rm AU}{50 D_\perp}\theta_\star \lesssim \pi_{\rm rel} \lesssim \kappa M_{\rm L} \left(\frac{\rm AU}{D_\perp}\right)^2\ .
\end{equation}
For terrestrial parallax ($D_\perp\sim R_\oplus$), the upper constraint on $\pi_{\rm rel}$ is almost always satisfied unless the lens is extremely low-mass or extremely nearby. The combined constraint thus suggests that only relatively nearby ($D_{\rm L}\lesssim 2.5$~kpc) lenses can be probed by terrestrial parallax, and since the lensing probability of such stars is quite low, the overall probability to measure isolated lens masses through this channel is extremely low. This has been shown in \citet{GouldYee:2013}, in which they found an event rate $\sim$1.6 Gyr$^{-1}$.
Observationally, in the era prior to the \emph{Spitzer} microlensing campaign, only two single-lens microlensing events yielded mass measurements \citep[OGLE-2007-BLG-224 \& OGLE-2008-BLG-279,][]{Gould:2009,Yee:2009} in this way, although more than $10,000$ events were discovered and at least four thousand of them were monitored as intensively as \emph{Spitzer} events.
\footnote{For example, among $\sim$10,000 microlensing events that were detected by OGLE-IV by 2015, more than 4,000 fall in the OGLE-IV high-cadence ($>10$ per night) fields. Those that reached high magnifications were followed up more intensively for the purpose of detecting planets \citep{GriestSafizadeh:1998}, and thus any deviation from the standard point-lens point-source microlensing light curve, such as the finite-source effect and terrestrial parallax, would be efficiently detected and further studied for the purpose of detecting planets.}

Equation~(\ref{eqn:pirellimits}) also shows that in a \textit{Spitzer}-like microlensing program ($D_\perp \sim 1~$AU), the parallax parameter is measurable for the majority of microlensing events. This has been proved by the 2014 \& 2015 \textit{Spitzer} microlensing programs \citep{CalchiNovati:2015a,CalchiNovati:2015b}. In addition, the lens-source relative trajectory is most often very different as seen from Earth and from the satellite. This means that the probability to detect the finite-source effect in single-lens events is almost doubled. Thus, the overall probability to make mass measurements of isolated microlenses for \textit{Spitzer} is $2\langle\rho\rangle/u_{0,\rm max}$, where $\langle\rho\rangle\sim0.005$ is the average source size normalized to the Einstein radius for \textit{Spitzer} events, and $u_{0,\rm max}\approx0.3$ is set as one of the criteria for selecting events \citep{Yee:2015b}. This theoretical estimate (3.3\%) agrees with the apparent frequency of such events ($3/170=1.8\%$) reasonably well, with the factor of $\sim$2 difference likely due to the incomplete coverage of \textit{Spitzer} data in some events. Hence, space-based microlensing experiments such as \textit{Spitzer} and \textit{Kepler} \citep{GouldHorne:2013,Henderson:2016} can significantly increase the probability to measure the mass of isolated microlenses, besides their primary goal of detections and statistical studies of planets. Furthermore, because the impact parameter of the observatory that shows the finite-source effect must be close to zero, the generic four-fold degeneracy in single lens cases is effectively broken \citep{GouldYee:2012}. 

Measuring the mass of isolated microlenses is of great scientific interest. \citet{Gould:2000} estimated that $\sim20\%$ of all Galactic microlensing events are caused by stellar remnants, and specifically that $\sim1\%$ are due to stellar mass black holes (BHs). In addition, there exists a population of probably unbound planet-mass objects that are almost twice as common as main-sequence stars \citep{Sumi:2011}. These isolated stellar remnants and free floating planets (FFPs) can only be probed by microlensing, and our work shows that space-based observations can significantly enhance the probability to definitely measure their masses. 

The future of detecting dark or extremely faint isolated objects via microlensing is promising. In addition to \emph{Spitzer}, the re-proposed \emph{Kepler} mission \citep[\emph{K2},][]{Howell:2014} is going to continuously monitor a 4 deg$^2$ microlensing fields for $\sim70$ days in its campaign 9, which provides so far a unique chance to probe especially the FFP population \citep{GouldHorne:2013,Henderson:2016}. Future space-based microlensing missions such as the Wide-Field InfraRed Survey Telescope \citep[\textit{WFIRST,}][]{Spergel:2015} and possibly Euclid \citep{Penny:2013}, once combined with a simultaneous ground-based telescope network, will significantly increase the sensitivities to especially FFPs with even lower masses \citep{ZhuGould:2016}.

\acknowledgements
Work by WZ, SCN and AG was supported by JPL grant 1500811. WZ and AG also acknowledge support by NSF grant AST-1516842.
Work by JCY was performed under contract with the California Institute of Technology (Caltech)/Jet Propulsion Laboratory (JPL) funded by NASA through the Sagan Fellowship Program executed by the NASA Exoplanet Science Institute.
The \emph{Spitzer} Team thanks Christopher S.\ Kochanek for graciously trading us his allocated observing time on the CTIO 1.3m during the \emph{Spitzer} campaign.
The OGLE project has received funding from the National Science Centre, Poland, grant MAESTRO 2014/14/A/ST9/00121 to AU. 
This work is based in part on observations made with the {\it Spitzer} Space Telescope, which is operated by the Jet Propulsion Laboratory, California Institute of Technology under a contract with NASA.
Work by C.H. was supported by Creative Research Initiative Program (2009-0081561) of National Research Foundation of Korea. This research has made the use of the KMTNet telescopes operated by the Korea Astronomy and Space Science Institute (KASI), and was supported by KASI grant 2016-1-832-01.
This work makes use of observations from the LCOGT network, which includes three SUPAscopes owned by the University of St Andrews. The RoboNet programme is an LCOGT Key Project using time allocations from the University of St Andrews, LCOGT and the University of Heidelberg together with time on the Liverpool Telescope through the Science and Technology Facilities Council (STFC), UK. This research has made use of the LCOGT Archive, which is operated by the California Institute of Technology, under contract with the Las Cumbres Observatory.
KH acknowledges support from STFC grant ST/M001296/1.
This project is partly supported by the Strategic Priority Research Program ``The Emergence of Cosmological Structures'' of the Chinese Academy of Sciences, Grant No. XDB09000000 (SM and SD).
M.P.G.H. acknowledges support from the Villum Foundation. 
N.P. acknowledges funding by the Gemini-Conicyt Fund, allocated to the project No. 32120036. 
Based on data collected by MiNDSTEp with the Danish 1.54 m telescope at the ESO La Silla observatory. Operation of the Danish 1.54m telescope at ESO’s La Silla observatory was supported by The Danish Council for Independent Research, Natural Sciences, and by Centre for Star and Planet Formation. The MiNDSTEp monitoring campaign is powered by ARTEMiS \citep[Automated Terrestrial Exoplanet Microlensing Search,][]{Dominik:2008}. D.M.B. acknowledges support from NPRP grant \# X-019-1-006 from the Qatar National Research Fund (a member of Qatar Foundation). GD acknowledges Regione Campania for support from POR-FSE Campania 2014-2020. O.W. and J.S. acknowledge support from the Communaut\'e francaise de Belgique $–$ Actions de recherche concert\'ees $–$ Acad\'emie universitaire Wallonie-Europe.

\end{CJK*}

\begin{thebibliography}{}
\bibitem[Bensby et al.(2013)]{Bensby:2013} Bensby, T., Yee, J.~C., Feltzing, S., et al.\ 2013, \aap, 549, A147
\bibitem[Bessell \& Brett(1988)]{BessellBrett:1988} Bessell, M.~S., \& Brett, J.~M.\ 1988, \pasp, 100, 1134
\bibitem[Boyajian et al.(2014)]{Boyajian:2014} Boyajian, T.~S., van Belle, G., \& von Braun, K.\ 2014, \aj, 147, 47
\bibitem[Calchi Novati et al.(2015a)]{CalchiNovati:2015a} Calchi Novati, S., Gould, A., Udalski, A., et al.\ 2015, \apj, 804, 20
\bibitem[Calchi Novati et al.(2015b)]{CalchiNovati:2015b} Calchi Novati, S., Gould, A., Yee, J.~C., et al.\ 2015, arXiv:1509.00037
\bibitem[Claret(2000)]{Claret:2000} Claret, A.\ 2000, \aap, 363, 1081
\bibitem[DePoy et al.(2003)]{Depoy:2003} DePoy, D.~L., Atwood, B., Belville, S.~R., et al.\ 2003, \procspie, 4841, 827
\bibitem[Diaz-Cordoves \& Gimenez(1992)]{DiazCordoves:1992} Diaz-Cordoves, J., \& Gimenez, A.\ 1992, \aap, 259, 227
\bibitem[Dominik et al.(2008)]{Dominik:2008} Dominik, M., Horne, K., Allan, A., et al.\ 2008, Astronomische Nachrichten, 329, 248
\bibitem[Dominik et al.(2010)]{Dominik:2010} Dominik, M., J{\o}rgensen, U.~G., Rattenbury, N.~J., et al.\ 2010, Astronomische Nachrichten, 331, 671
\bibitem[Dong et al.(2007)]{Dong:2007} Dong, S., Udalski, A., Gould, A., et al.\ 2007, \apj, 664, 862
\bibitem[Fields et al.(2003)]{Fields:2003} Fields, D.~L., Albrow, M.~D., An, J., et al.\ 2003, \apj, 596, 1305
\bibitem[Foreman-Mackey et al.(2013)]{ForemanMackey:2013} Foreman-Mackey, D., Hogg, D.~W., Lang, D., \& Goodman, J.\ 2013, \pasp, 125, 306
\bibitem[Gaudi \& Gould(1997)]{GaudiGould:1997} Gaudi, B.~S., \& Gould, A.\ 1997, \apj, 486, 85
\bibitem[Gould(1992)]{Gould:1992} Gould, A.\ 1992, \apj, 392, 442
\bibitem[Gould(1994)]{Gould:1994a} Gould, A.\ 1994, \apjl, 421, L71 
\bibitem[Gould(1994)]{Gould:1994b} Gould, A.\ 1994, \apjl, 421, L75
\bibitem[Gould(1997)]{Gould:1997} Gould, A.\ 1997, \apj, 480, 188 
\bibitem[Gould(2000)]{Gould:2000} Gould, A.\ 2000, \apj, 535, 928
\bibitem[Gould(2013)]{Gould:2013} Gould, A.\ 2013, \apjl, 763, L35
\bibitem[Gould \& Loeb(1992)]{GouldLoeb:1992} Gould, A., \& Loeb, A.\ 1992, \apj, 396, 104
\bibitem[Gould \& Welch(1996)]{GouldWelch:1996} Gould, A., \& Welch, D.~L.\ 1996, \apj, 464, 212
\bibitem[Gould \& Yee(2012)]{GouldYee:2012} Gould, A., \& Yee, J.~C.\ 2012, \apjl, 755, L17
\bibitem[Gould \& Horne(2013)]{GouldHorne:2013} Gould, A., \& Horne, K.\ 2013, \apjl, 779, L28
\bibitem[Gould \& Yee(2013)]{GouldYee:2013} Gould, A., \& Yee, J.~C.\ 2013, \apj, 764, 107
\bibitem[Gould et al.(2009)]{Gould:2009} Gould, A., Udalski, A., Monard, B., et al.\ 2009, \apjl, 698, L147
\bibitem[Gould et al.(2010)]{Gould:2010} Gould, A., Dong, S., Gaudi, B.~S., et al.\ 2010, \apj, 720, 1073
\bibitem[Gould et al.(2014)]{Gould:2014} Gould, A., Carey, S., \& Yee, J.\ 2014, Spitzer Proposal, 11006
\bibitem[Griest \& Safizadeh(1998)]{GriestSafizadeh:1998} Griest, K., \& Safizadeh, N.\ 1998, \apj, 500, 37
\bibitem[Henderson et al.(2015)]{Henderson:2016} Henderson, C.~B., Penny, M., Street, R.~A., et al.\ 2015, arXiv:1512.09142
\bibitem[Honma(1999)]{Honma:1999} Honma, M.\ 1999, \apjl, 517, L35
\bibitem[Howell et al.(2014)]{Howell:2014} Howell, S.~B., Sobeck, C., Haas, M., et al.\ 2014, \pasp, 126, 398
\bibitem[Kim et al.(2016)]{Kim:2016} Kim, S.-L., Lee, C.-U., Park, B.-G., et al. 2016, JKAS, 49, 37
\bibitem[Kervella et al.(2004)]{Kervella:2004} Kervella, P., Th{\'e}venin, F., Di Folco, E., \& S{\'e}gransan, D.\ 2004, \aap, 426, 297
\bibitem[Kipping(2013)]{Kipping:2013} Kipping, D.~M.\ 2013, \mnras, 435, 2152
\bibitem[Mao \& Paczynski(1991)]{MaoPaczynski:1991} Mao, S., \& Paczynski, B.\ 1991, \apjl, 374, L37
\bibitem[Nataf et al.(2013)]{Nataf:2013} Nataf, D.~M., Gould, A., Fouqu{\'e}, P., et al.\ 2013, \apj, 769, 88
\bibitem[Paczynski(1986)]{Paczynski:1986} Paczynski, B.\ 1986, \apj, 304, 1
\bibitem[Penny et al.(2013)]{Penny:2013} Penny, M.~T., Kerins, E., Rattenbury, N., et al.\ 2013, \mnras, 434, 2 
\bibitem[Poleski et al.(2014)]{Poleski:2014} Poleski, R., Skowron, J., Udalski, A., et al.\ 2014, \apj, 795, 42
\bibitem[Refsdal(1966)]{Refsdal:1966} Refsdal, S.\ 1966, \mnras, 134, 315
\bibitem[Saito et al.(2012)]{Saito:2012} Saito, R.~K., Hempel, M., Minniti, D., et al.\ 2012, \aap, 537, A107
\bibitem[Shvartzvald et al.(2015)]{Shvartzvald:2015} Shvartzvald, Y., Udalski, A., Gould, A., et al.\ 2015, arXiv:1508.06636
\bibitem[Smith et al.(2007)]{Smith:2007} Smith, M.~C., Wo{\'z}niak, P., Mao, S., \& Sumi, T.\ 2007, \mnras, 380, 805
\bibitem[Spergel et al.(2015)]{Spergel:2015} Spergel, D., Gehrels, N., Baltay, C., et al.\ 2015, arXiv:1503.03757 
\bibitem[Street et al.(2015)]{Street:2015} Street, R.~A., Udalski, A., Calchi Novati, S., et al.\ 2015, arXiv:1508.07027
\bibitem[Sumi et al.(2011)]{Sumi:2011} Sumi, T., Kamiya, K., Bennett, D.~P., et al.\ 2011, \nat, 473, 349
\bibitem[Tsapras et al.(2009)]{Tsapras:2009} Tsapras, Y., Street, R., Horne, K., et al.\ 2009, Astronomische Nachrichten, 330, 4
\bibitem[Udalski et al.(1994)]{Udalski:1994} Udalski, A., Szymanski, M., Stanek, K.~Z., et al.\ 1994, \actaa, 44, 165
\bibitem[Udalski(2003)]{Udalski:2003} Udalski, A.\ 2003, \actaa, 53, 291 
\bibitem[Udalski et al.(2015)]{Udalski:2015a} Udalski, A., Yee, J.~C., Gould, A., et al.\ 2015, \apj, 799, 237
\bibitem[Udalski et al.(2015)]{Udalski:2015b} Udalski, A., Szyma{\'n}ski, M.~K., \& Szyma{\'n}ski, G.\ 2015, \actaa, 65, 1 
\bibitem[Witt \& Mao(1994)]{WittMao:1994} Witt, H.~J., \& Mao, S.\ 1994, \apj, 430, 505 
\bibitem[Yee et al.(2009)]{Yee:2009} Yee, J.~C., Udalski, A., Sumi, T., et al.\ 2009, \apj, 703, 2082
\bibitem[Yee(2015)]{Yee:2015c} Yee, J.~C.\ 2015, arXiv:1509.05043
\bibitem[Yee et al.(2015a)]{Yee:2015a} Yee, J.~C., Udalski, A., Calchi Novati, S., et al.\ 2015, \apj, 802, 76
\bibitem[Yee et al.(2015b)]{Yee:2015b} Yee, J.~C., Gould, A., Beichman, C., et al.\ 2015, \apj, 810, 155
\bibitem[Yoo et al.(2004)]{Yoo:2004} Yoo, J., DePoy, D.~L., Gal-Yam, A., et al.\ 2004, \apj, 603, 139
\bibitem[Zhu et al.(2015a)]{Zhu:2015a} Zhu, W., Udalski, A., Gould, A., et al.\ 2015, \apj, 805, 8 
\bibitem[Zhu et al.(2015b)]{Zhu:2015b} Zhu, W., Gould, A., Beichman, C., et al.\ 2015, \apj, 814, 129 
\bibitem[Zhu \& Gould(2016)]{ZhuGould:2016} Zhu, W., \& Gould, A.\ 2016, arXiv:1601.03043 
\end{thebibliography}
\end{document}